\documentclass[prl,twocolumn,showpacs,preprintnumbers,superscriptaddress]{revtex4}

\usepackage{times}
\usepackage{subfig}
\usepackage{bm}
\usepackage{graphicx}
\usepackage{amsbsy}
\usepackage{amsmath}
\usepackage{amsfonts}
\usepackage{amsthm}
\usepackage{float}

\begin{document}

\theoremstyle{plain}
\newtheorem{theorem}{Theorem}
\newtheorem{lemma}[theorem]{Lemma}
\newtheorem{corollary}[theorem]{Corollary}
\newtheorem{proposition}[theorem]{Proposition}
\newtheorem{conjecture}[theorem]{Conjecture}

\theoremstyle{definition}
\newtheorem{definition}[theorem]{Definition}

\title{\bf Quantum Correlations in Quantum Cloning }
\author {Indranil Chakrabarty}
\affiliation{Institute of Physics, Sainik School Post, 
Bhubaneswar-751005, Orissa, India}
\date{}

\begin{abstract}
We utilize quantum discord to charecterize the correlation present in Buzek-Hillery quantum copying machine \cite{bh} (not necessarily universal
quantum cloning machine). In other words we quantify the correlation present beetween the original and the replicated copy
of the quantum state obtained at the outport port,  Interestingly, we find some domain of the machine parameter, for which the 
quantum disord is non negative even in the mere absence of entanglement. These non zero values of the quantum discord is a strong signature 
for the presence of non classical correlations. This is one step forward evidence in the support of the fact that quantum discord and entanglement 
are not synonymous.  \\
\end{abstract}

\maketitle

\section{Introduction } Recently there have been a lot of discussion regarding the role of entanglement in explaining  all possible non classical correlation present in a bipartite  quantum system \cite{oll,hen,luo,ur}. The most interesting example in this context is the Werner state. Few years back, Ollivier and Zurek devoloped a new way of charecterizing the non classical correlations and they named this measure as "Quantum Discord" \cite{oll}. This particlar measure was sucessful in quantifying the non classical correlation in Werner which was not detected by the separability criterion. Later it was also shown that discord charecterize the correlations present in various mixed state quantum computational model like  DQC1 model (deterministic quantum computation with one bit) \cite{ani}. Recently a group of authors presented an alternative way of quantifying the correlation in multipartie system and they named it as "Quantum Dissonance" \cite{mw10}. In another work we also study the role of the quantum discord in open quantum systems.
The states studied are modelled by two-qubit system interacting with its environment via quantum nondemolition (purely dephasing) as well as dissipative type of interaction \cite{ind}. Recently we introduce the concept of 'Quantum Dissension', which is a kind of generalization of quantum discord in the multipartite system \cite{ind1}.

Another interesting feature in the quantum information theory is the quantum cloning. Most of us are well known of 'No Cloning ' theorem \cite{woo}. Although nature prevents us from amplifying an unknown quantum state but in principle it is always possible to construct a quantum cloning machine that replicates an unknown quantum state approximately \cite{bh,ind2,gis}. Quantum copying machine can be mainly categorized into two classes: (a) Deterministic quantum copying machine and (b) Probabilistic quantum copying machine. The first type of quantum cloning machine can be of two types: (i) State dependent quantum cloning machine, for example Wootters-Zurek (W-Z) quantum cloning machine whose copying quality depends on the input state \cite{woo}, (ii) Universal quantum copying machine, for example Buzek-Hillery (B-H) quantum cloning machine \cite{bh}, whose copying quality remains same for all input state. In addition, the performance of universal B-H quantum cloning machine is, on average, better than that of the state dependent W-Z cloning machine. The fidelity of cloning of B-H universal quantum copying machine is $\frac{5}{6}$ which is better than any other existing universal quantum cloning machine. The Probabilistic quantum cloning machine clones an unknown quantum state, secretly chosen from a certain set of linearly independent states, accurately but with certain probability less than unity \cite{dua,dua1}.  At this point one might ask an important question that whether quantum correlation is responsible for cloning or not. Initially there is no such correlation between the input state and the blank state. This is because they are the individual systems. However at the output port we always obtain a combined state, which is most of the times correlated. It is not clear that whether this correlation is entanglement or something else.

In order to find an answer to this question, here in this work we consider a particular type of cloning machine; B-H cloning machine (not necessarily universal quantum cloning machine) and hence try to quantify the amount of correlation present in the mixed two qubit output state.
We charecterize this quantum correlation with quantum disord and see that this is not equivalent to what we actually mean by the term entanglement. In support of this argument we find out some domain of the machine parameter of the 
cloning machine for which the output state have non vanishing discord inspite of being separable.  
\section{Analysis of the output of Buzek-Hillery copying machine}
In this section we study the correlation present in the  output
copies of Buzek-Hillery cloning machine  \cite{bh}. But before that we give a short description of the Buzek-Hillery cloning machine.  \\
The action of the Buzek-Hillery quantum cloning machine \cite{bh} is
given by
\begin{eqnarray}
|0\rangle_{a}|0\rangle_{b}|Q\rangle_{x}\longrightarrow|0\rangle_{a}|0\rangle_{b}|Q_0\rangle_{x}
+[|0\rangle_a|1\rangle_b+|1\rangle_a|0\rangle_b]|Y_0\rangle_x\\
|1\rangle_{a}|0\rangle_{b}|Q\rangle_{x}\longrightarrow|1\rangle_{a}|1\rangle_{b}|Q_1\rangle_{x}
+[|0\rangle_a|1\rangle_b+|1\rangle_a|0\rangle_b]|Y_1\rangle_x
\end{eqnarray}
The unitarity and the orthogonality of the cloning transformation demands,
\begin{eqnarray}
_{x}\langle Q_i|Q_i\rangle_{x}+2_{x}\langle Y_i|Y_i\rangle_{x}
=1~~~~(i=0,1)\\
 _{x}\langle Y_0|Y_1\rangle_{x}=_{x}\langle
Y_1|Y_0\rangle_{x}=0
\end{eqnarray}
Here the copying machine state vectors $|Y_i\rangle_{x}$ and
$|Q_i\rangle_{x}$ are assumed to be mutually orthogonal, so are
the state vectors $\{|Q_0\rangle,|Q_1\rangle \}$.\\
Let us consider a quantum state which is to be cloned
\begin{eqnarray}
|\chi\rangle= \alpha|0\rangle+\beta|1\rangle
\end{eqnarray}
where $\alpha^{2}+\beta^{2}=1$.\\
Here we confine ourselves to a limited class of input states (5),
where $\alpha$ and $\beta$ are real.\\
After using the cloning transformation (1-2) on quantum state (5)
and tracing out the machine state vector, the two qubit reduced
density operator describing the two clones is given by
\begin{eqnarray}
\rho_{ab}^{out}=&&\alpha^2(1-2j)|00\rangle\langle
00|+\frac{\alpha\beta}{\sqrt{2}}(1-2j)|00\rangle\langle
+|{}\nonumber\\&&+\frac{\alpha\beta}{\sqrt{2}}(1-2j)|+\rangle\langle
00|+2j|+\rangle\langle
+|{}\nonumber\\&&+\frac{\alpha\beta}{\sqrt{2}}(1-2j)|+\rangle\langle 11|
+\frac{\alpha\beta}{\sqrt{2}}(1-2j)|11\rangle\langle
+|{}\nonumber\\&&+\beta^2(1-2j)|11\rangle\langle 11|
\end{eqnarray}
where we have used the following notations\\
$_{x}\langle Y_0|Y_0\rangle_{x}$=$_{x}\langle Y_1|Y_1\rangle_{x}$=$j$,\\
$_{x}\langle Y_0 |Q_1\rangle_{x}$=$_{x}\langle Q_0
|Y_1\rangle_{x}$=$_{x}\langle Q_1|Y_0\rangle_{x}$=$_{x}\langle Y_1
|Q_0\rangle_{x}$=$ \
\frac{n}{2}$,\\
$|+\rangle=\frac{1}{\sqrt{2}}(|01\rangle+|10\rangle)$ (where $n=1-2j$, $j$ is the machine parameter). The output state $\rho_{ab}^{out}$
is of our prime importance. Before quantifying the amount of correlation present in it in terms 
of quantum discord, we give a basic idea of quantum discord. 

In classical information theory \cite{cover}, we define  mutual information between two random variables
$X$ and $Y$ as $I(X:Y)=H(X)-H(X|Y)$, where $H(X),H(X|Y)$ are the entropy of $X$ and the conditional entropy of $X$ given that $Y$ has 
already occurred. Mutual information actually quantifies the amount of reduction in the uncertainty 
about one random variable because of the occurrence of the other random variable. Since 
$H(X|Y)=H(X,Y)-H(Y)$, so one can equivalently write the expression of mutual information as $J(X:Y)=H(X)+H(Y)-H(X,Y)$. 
Classically, these two expressions are identical. However, when we consider the same expressions 
in the quantum domain the random variables $X,Y$  are replaced by  the density matrices $\rho_X, \rho_Y$ and the Shannon entropies $H(X), H(Y)$ by Von-neumann entropies (e.g: $H(X)=H(\rho_X)=-{\rm Tr}[\rho_X\log(\rho_X)$]). With these
replacements, we can have an expression for $J$ in the quantum case. However inorder to obtain an analogous expression for $I$, one needs to specify
the conditional entropy $H(X|Y)$. Now, from the definition itself the conditional entropy $H(X|Y)$
requires a specification of the state of $X$ given the state of $Y$.  However in quantum theory, there is no scope of making such a system  until the 
to-be-measured set of states of the system  $Y$ are selected. For that reason we focus on perfect measurements of Y defined by a set of one dimensional
projectors $\{\pi^Y_j\}$. The index $j$ represents 
different outcomes obtained as a result of this measurement.
The state of X, after the measurement is given by
\begin{eqnarray}
\rho_{X|\pi^Y_j}=\frac{\pi^Y_j \rho_{XY} \pi^Y_j}{Tr(\pi^Y_j \rho_{XY})},
\end{eqnarray}
The probability of obtaining this measurement outcome is given by $p_j={Tr(\pi^Y_j \rho_{XY})}$. Thus, $H(\rho_{X|\pi^Y_j})$ 
gives us  Von-neumann entropy of the state $\rho_X$, provided that the projective measurement 
is carried out on the system $Y$ in the most general basis $\{\cos(t)|0\rangle+\sin(t)|1\rangle,\sin(t)|0\rangle-\cos(t)|1\rangle\}$ (where $t$ is the azimuthal angle). 
 
The entropies $H(\rho_{X|\pi^Y_j})$ weighted by the probabilities $p_j$, yield  the conditional entropy
of X,  given the complete set of measurements $\{\pi^Y_j\}$
 on Y,  as $H(X|\{\pi^Y_j\})=\sum_j p_jH(\rho_{X|\pi^Y_j})$.  From this, the quantum analogue of $I(X:Y)$ is seen to be 
\begin{eqnarray}
I(X:Y)=H(X)-H(X|\{\pi^Y_j\}),
\end{eqnarray}
while $J(X:Y)$ is similar to its classical counterpart
\begin{eqnarray}
J(X:Y)=H(X)+H(Y)-H(X,Y).
\end{eqnarray}
It is clearly evident that these two
expressions are not identical in standard quantum theory. The quantum discord is
this difference, 
\begin{eqnarray}
D(X:Y)=H(Y)-H(X,Y)+H(X|\{\pi^Y_j\}).
\end{eqnarray}
This is to be minimized over 
all sets of one dimensional projectors $\{\pi^Y_i\}$.    

\begin{figure}[tbp]
\begin{center}
{\includegraphics[height=4cm,width=4cm]{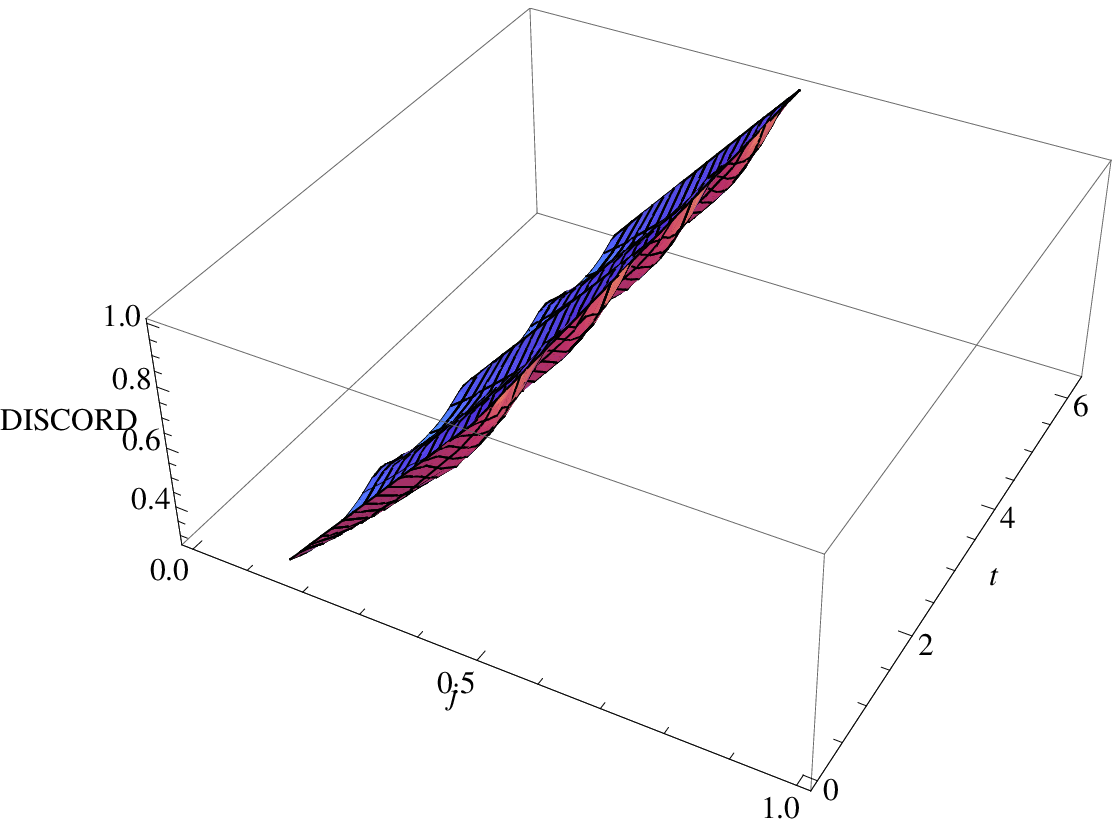}}
{\includegraphics[height=4cm,width=4cm]{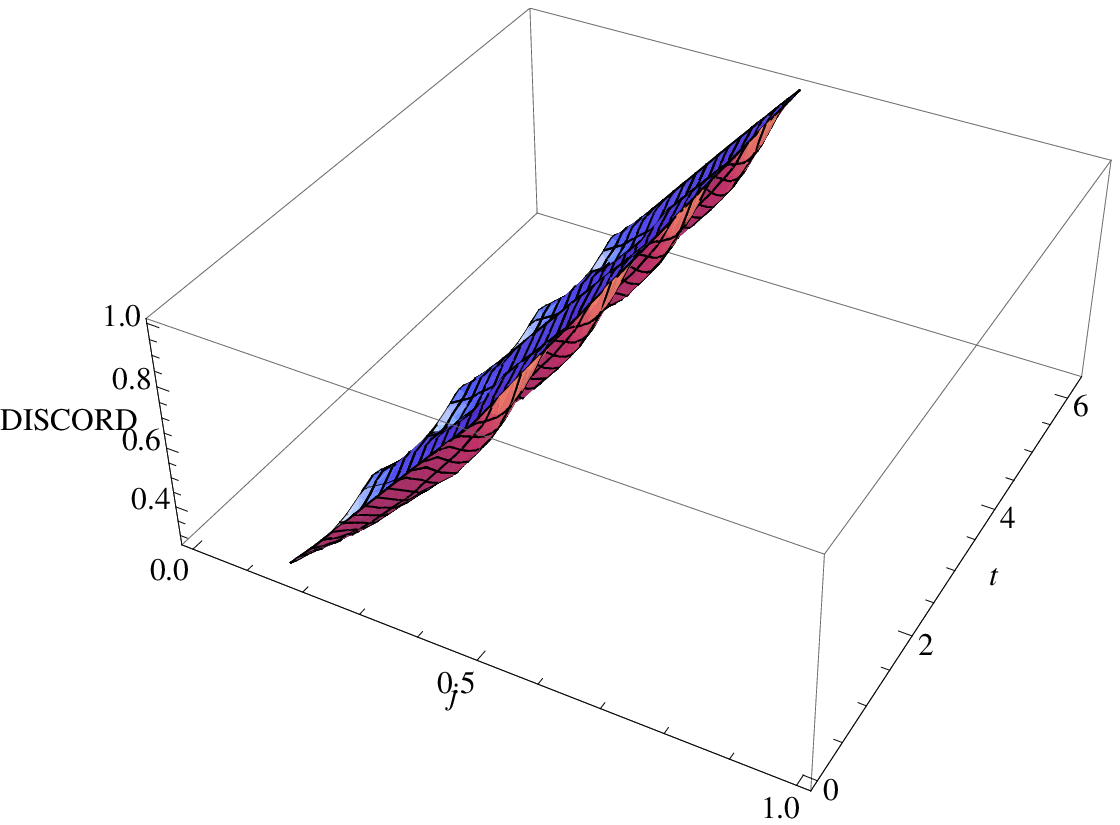}}
{\includegraphics[height=4cm,width=4cm]{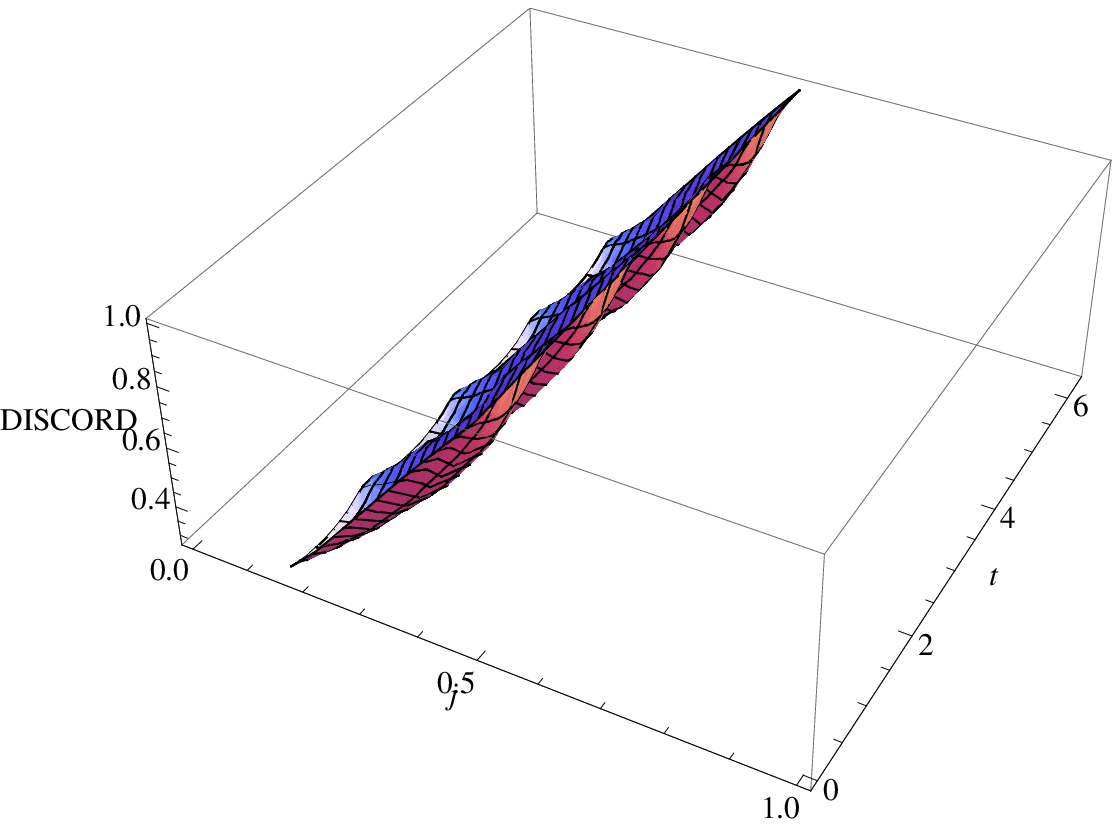}}
{\includegraphics[height=4cm,width=4cm]{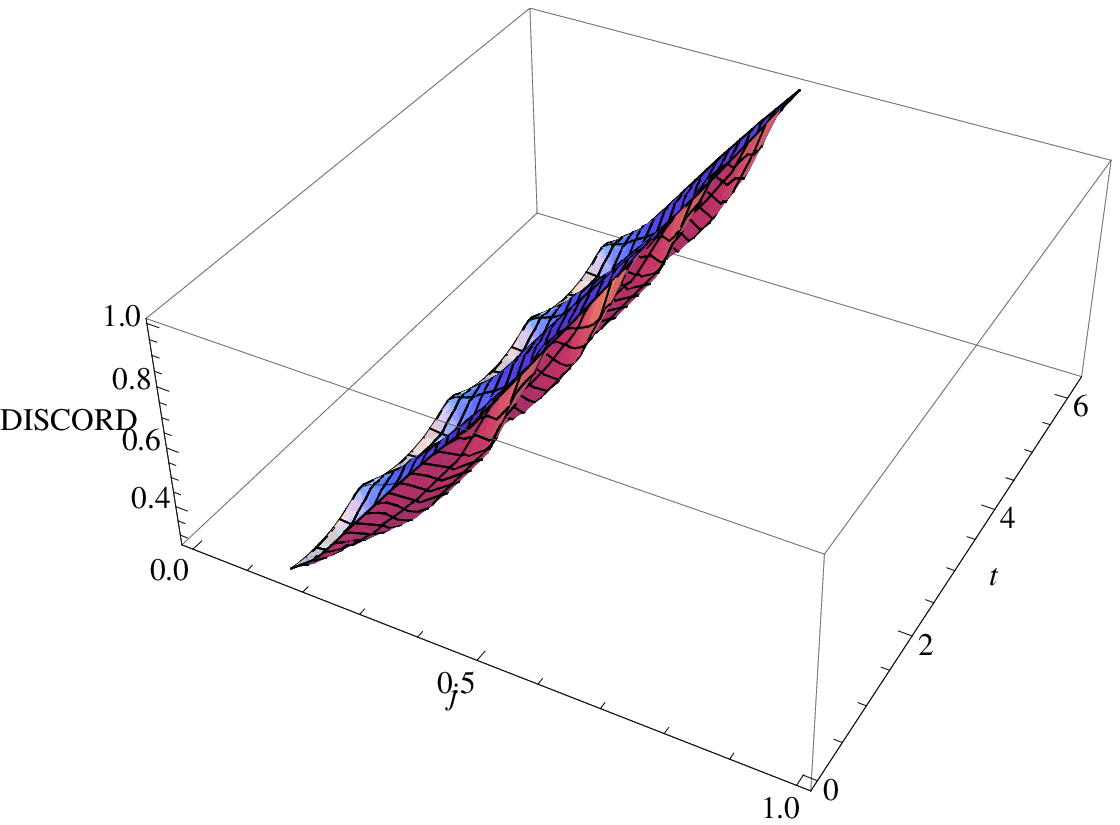}}
{\includegraphics[height=4cm,width=4cm]{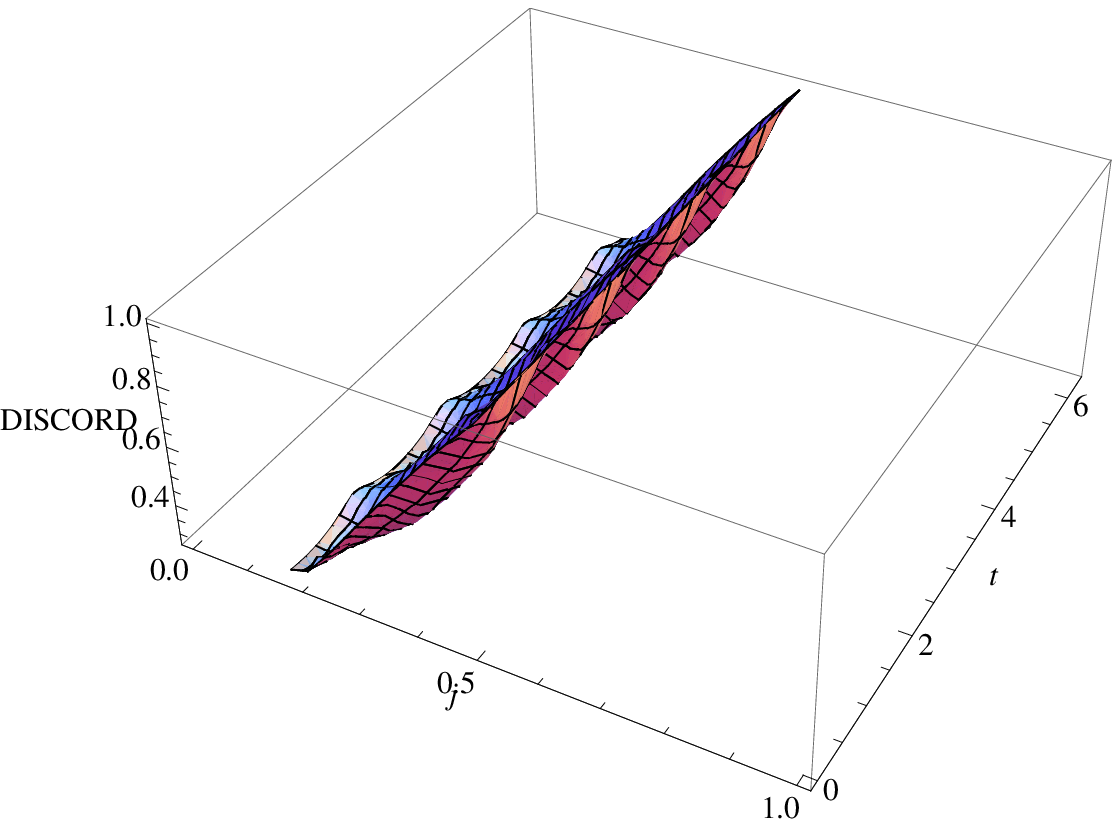}}
{\includegraphics[height=4cm,width=4cm]{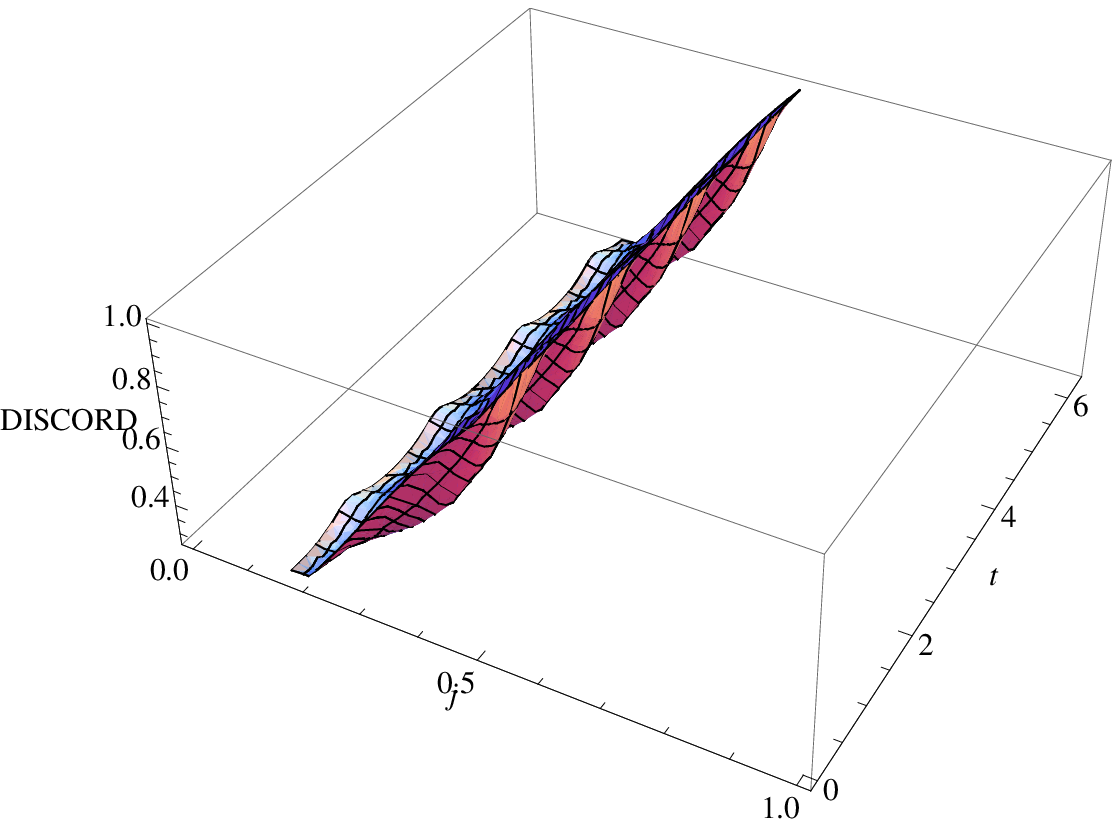}}
{\includegraphics[height=4cm,width=4cm]{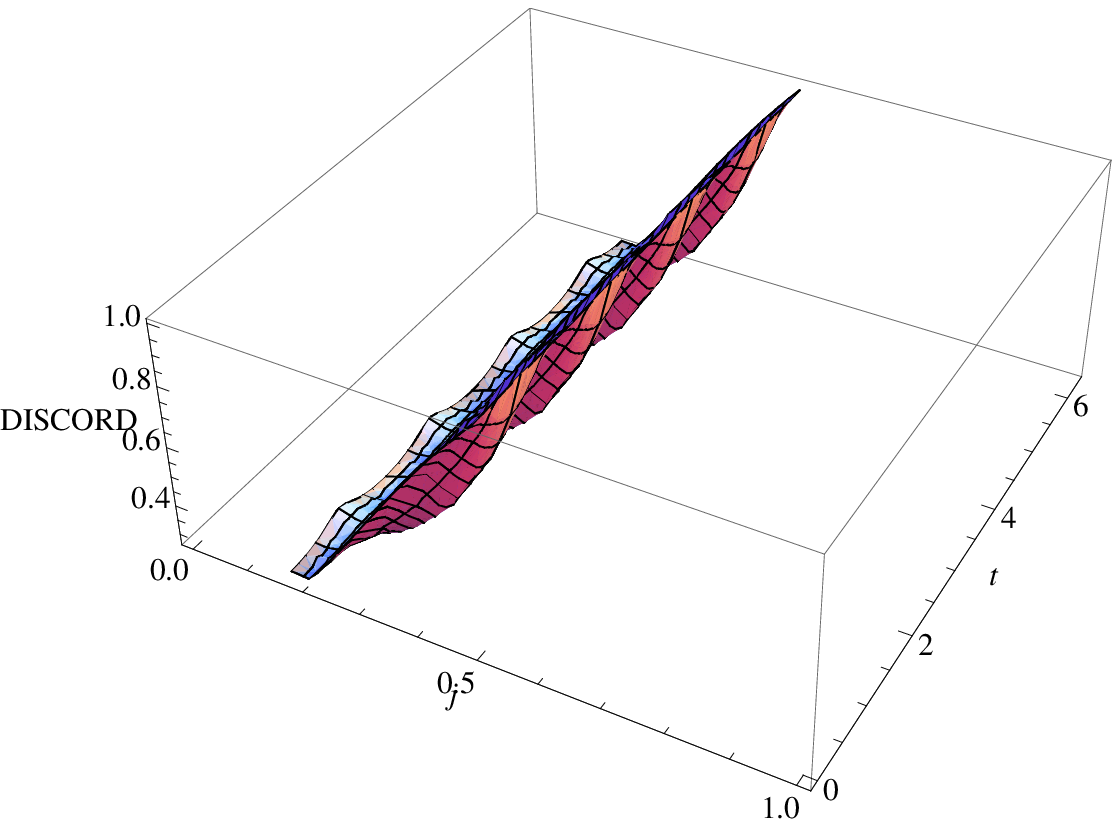}}
{\includegraphics[height=4cm,width=4cm]{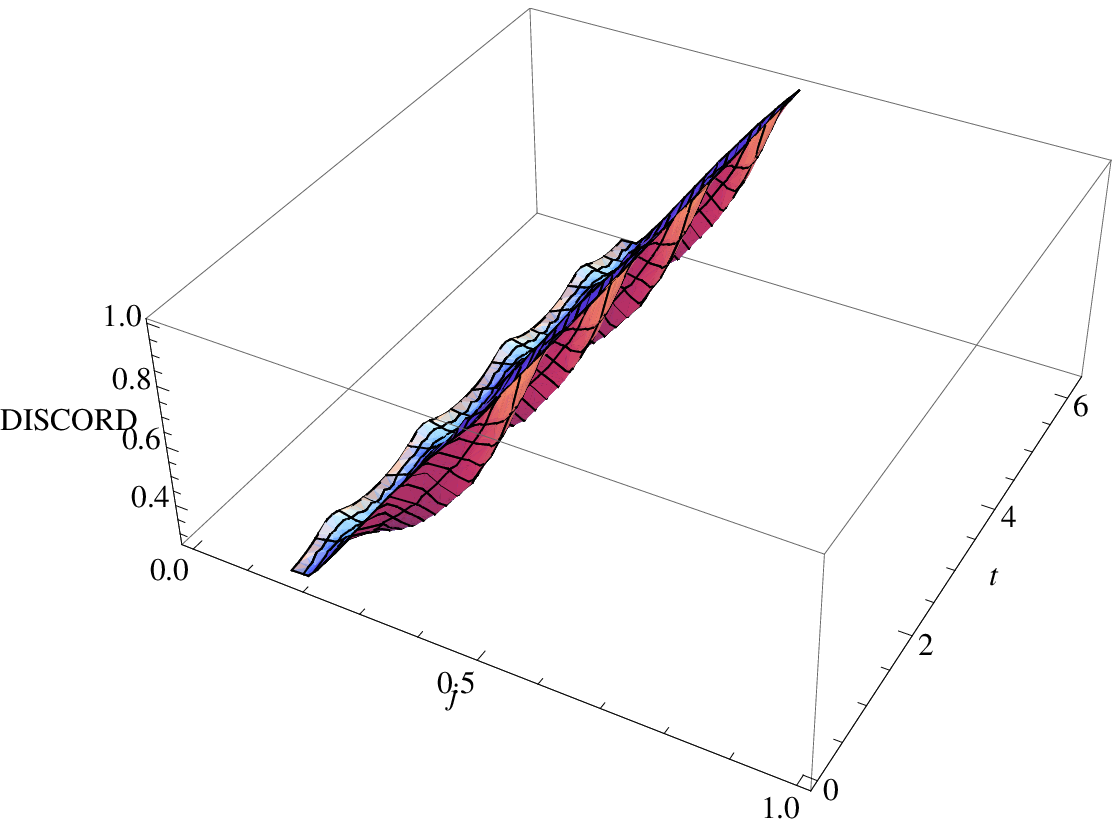}}
{\includegraphics[height=4cm,width=4cm]{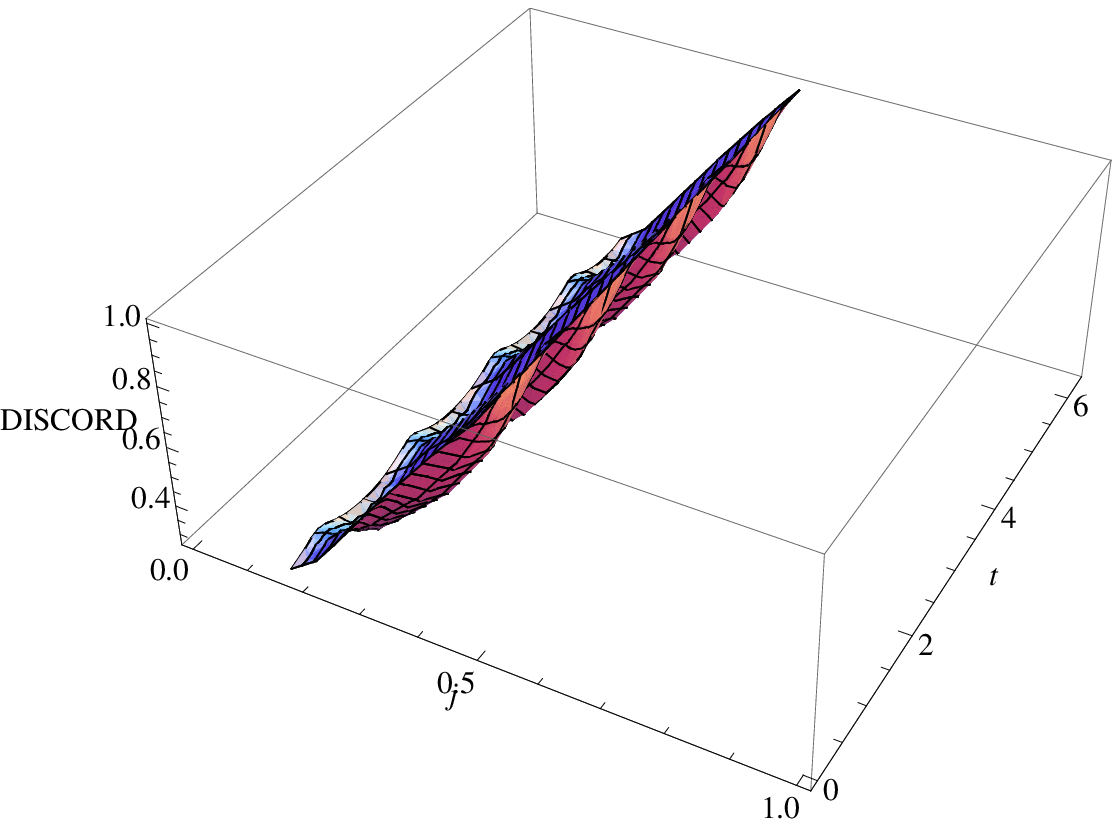}}
\end{center}
\caption{{\protect\small Quantum discord for  two-qubit system $\rho_{ab}^{out}$ obtained as a output
of Buzek-Hillery quantum cloning machine.  The Figures from the top left to the bottom center (left to right, then top to bottom) represent the quantum discord as a 
function of the machine parameter $j$ and the azimuthal angle $t$ for $\alpha=0.1,0.2,....0.9$ respectively \cite{thi}.}}\label{fig1}
\end{figure}

Here in this work we try to obtain the correlation present in the output state $\rho_{ab}^{out}$ in terms of the quantum discord. So far we have seen how discord can be used to characterize the nonclassical nature of the correlations in quantum states.
We now apply these ideas to the Buzek Hillery quantum cloning machine. We consider the output state $\rho_{ab}^{out}$ and consequently
find out the joint Von-neumann entropy as well as the Von-neumann entropy of the reduced subsystem $\rho_{b}^{out}$. Inorder to obtain the conditional entropy, projective measurement is carried 
out on the subsystem $b$ in the most general basis $\{\cos(t)|0\rangle+\sin(t)|1\rangle,\sin(t)|0\rangle-\cos(t)|1\rangle\}$ . On substuting these values of the entropies in the expression of discord we obtain discord as a function of the machine parameter $j$ and the azimuthal angle $t$ for different values 
of the input parameter $\alpha=0.1,0.2,.....0.9$. These are well demonstrated in the Figure \ref{fig1}. If we analyze each of these figures we obtain the non zero values of the discord for all values of $j$. This clearly indicates the presence of non classical correlation independent of the machine parameter.

The next question that comes forward whether this non classical correlation that we obtain in form of discord is the same thing 
as the entanglement or not. Indeed it is not so. Inorder to find out whether the output state is entangled or not, we take the help 
of the Peres Horodecki criterirea \cite{peres,3h}. It tells us that the necessary and sufficient condition for a state $\rho$  to be
inseparable is that at least one of the eigen values of the
partially transposed operator defined as
$\rho^{T}_{m\mu,n\nu}=\rho_{m\nu,n\mu}$ (where the Latin indices refer to the first subsystem, Greek indices to the second one ) is negative
\cite{peres,3h}.
This is equivalent to the condition that at least one of the two determinants

$W_{3}= \begin{array}{|ccc|}
  \rho_{00,00} & \rho_{01,00} & \rho_{00,10} \\
  \rho_{00,01} & \rho_{01,01} & \rho_{00,11} \\
  \rho_{10,00} & \rho_{11,00} & \rho_{10,10}
\end{array}$ and $W_{4}=\begin{array}{|cccc|}
   \rho_{00,00} & \rho_{01,00} & \rho_{00,10} & \rho_{01,10}\\
  \rho_{00,01} & \rho_{01,01} & \rho_{00,11} & \rho_{01,11} \\
  \rho_{10,00} & \rho_{11,00} & \rho_{10,10} & \rho_{11,10} \\
  \rho_{10,01} & \rho_{11,01} & \rho_{10,11} & \rho_{11,11}
\end{array}$
is negative.

Now we investigate the separability and inseparability of two qubit density
operator $\rho^{out}_{ab}$ for different discrete values of the input parameter  $\alpha$.\\
For the density matrix $\rho^{out}_{ab}$, we calculate the
determinants $W_{3}$ and $W_{4}$, which are given by
\begin{eqnarray}
W_{3}= \frac{\alpha^{2}j(1-2j)}{2}[2j-\beta^{2}(1-2j)],\\
W_{4}=\frac{1}{2}[\alpha^{2}\beta^{2}j(1-2j)^{2}(6j-1)-2j^{4}]
\end{eqnarray}
We find out the range of the machine parameter $j$ for different tabulated values of $\alpha$ and consequently 
check the separbility and inseparability of the output density matrix.


{\bf TABLE 1:}

\begin{tabular}{|c|c|c|c|}
\hline ($\alpha$)  & $W_3\geq 0, W_4\geq 0$ (range of $j\in(0,1)$) & Remark \\
\hline 0.1 & $j<0$ &  Inseparable \\
\hline 0.2 & $j<0$ & Inseparable \\
\hline 0.3 & $j<0$ & Inseparable \\
\hline 0.4 & $j<0$ & Inseparable \\
\hline 0.5 & $j<0$ & Inseparable \\
\hline 0.6 & $j\in [0.196,0.238]$ & Separable \\
\hline 0.7 & $j\in [0.191,0.250]$ & Separable \\
\hline 0.8 & $j\in [0.196,0.238]$ & Separable \\
\hline 0.9 & $j<0$ & Inseparable \\
\hline
\end{tabular}

It is clearly evident from the above table that for $\alpha=0.1,0.2,0.3,0.4,0.5,0.9$, there is no value of $j$
for which the output states are separable. So we conclude that for these values of $\alpha$, the output states are entangled.
On the other hand we find that for $\alpha=0.6,0.7,0.8$, we have a certain range of values of $j\in [0.196,0.238], [0.191,0.250],[0.196,0.238]$ (respectively) for which the output states are separable.

Thus we see that the quantum discord and the entanglement are not even equivalent features. Quantum discord comes out strongly as a better measure for non classical correlation. This is because of its ability to quantify the quantum correlation in those ranges of $j$ for which the entanglement is missing.

\section{Conclusions }
In a nutshell, here in this work we try to charecterize the non classical correlation which seems to be responsible for cloning. In that
context, we consider the Buzek-Hillery quantum cloning machine \cite{bh} and analyzed the correlation both in terms of quantum discord and also with the inseparability criterion. In fact we find out some range of the value of the machine parameter $j$ for given values $\alpha$, for which the output state is separable inspite of having non zero values of the quantum discord. Thus we see that both inseparability and discord are two individually different features of 
understanding the non classical correlation; the latter one proving to be better than the former one.  Thus we propose quantum discord to be a quantity which provides a best possible way of characterizing the performance of quantum cloning machine at least in the context of non classical correlations.

\end{document}